\documentclass[12pt]{iopart}

\usepackage{graphicx}

\newcommand{\GE}{gradient expansion}

\newcommand{\TF}{Thomas--Fermi}
\newcommand{\HF}{Hartree--Fock}

\newcommand{\Wz}{Weizs\"acker}

\newcommand{\eeref}[2]{equations~(\ref{#1}) and (\ref{#2})}

\usepackage{soul} 
\usepackage{color}
\usepackage{caption2}

\newcommand{\fig}[3]{
\begin{figure}[htbp]\centering\colorbox{paleyellow}
{\includegraphics[width=0.92\linewidth]{#3}}
\caption{#1}\label{#2}
\end{figure}
}
\captionstyle{normal}
\definecolor{darkblue}{rgb}{0,0,0.6}
\definecolor{paleyellow}{rgb}{1,1,1.0}
\newlength{\figlength} \figlength=0.9\textwidth

\setcaptionmargin{0.03in} 

\begin{document}

\title[Correction to kinetic energy density using exactly solvable model]{Correction to kinetic energy density using exactly solvable model}

\author{Alexey Sergeev, Raka Jovanovic, Sabre Kais and Fahhad H Alharbi}

\address{Qatar Environment and Energy Research Institute (QEERI), PO Box 5825, Doha, Qatar}

\ead{falharbi@qf.org.qa}
\vspace{10pt}
\begin{indented}
\item[]April 2015
\end{indented}

\begin{abstract}
An accurate non-gradient-expansion based correction to \TF\ is developed using solvable model.  The used model is  a system of $N$ non-interacting electrons moving independently in the Coulomb field of the nuclear charge. The presented correction is applicable for atoms and should be extendable beyond that. The method exploits the fact that the difference between the \TF\ approximation and the non-interacting kinetic energy is comparable to the difference between the same values inside the proposed solvable model. The numerical experiments show that by adding this correction factor, the precision of \TF\ approximation is enhanced by an order of magnitude.
\end{abstract}

%
%
%
%
%
\section{Introduction}
As known, the original density functional theory (DFT) is based on Hohenberg and Kohn prominent work \cite{WOS:A19641557C00018} where they proved that the ground state of any many-electron system is completely characterized by its density and that the energy functional of the system attains its minimum at the density corresponding to the ground state. However, representing the contribution of kinetic energy as density functional has proven to be challenging as the accuracy and applicability of the proposed kinetic energy density functionals (KEDF) are generally not sufficient \cite{B01,X01,M01,ribeiro2015}. Alternatively, Kohn and Sham (KS-DFT) \cite{WOS:A19657000000015} suggested an approach where the``orbitals'' are reintroduced where the sum of the orbitals' densities equals to the density of the real system and the kinetic energy is defined as the kinetic energy of the introduced ``fictitious'' system. Computational-wise, this results in converting the problem back from 3-dimensional (3D) to 3$N$ dimension  as $N$ orbitals are determined by solving the governing $N$ equations self-consistently \cite{WOS:A19657000000015,B02,M02}, where $N$ is the number of the particles. Despite this drawback, KS-DFT nowadays dominates atomistic calculations \cite{B02}. In parallel, the work to find an accurate KEDF is still active, but modestly. In the recent years, it starts gaining attention \cite{K01,S01,K02,ribeiro2015}. To distinguish this DFT doctrine from KS-DFT which depends on the orbital, it is commonly named ``orbital-free' DFT (OF-DFT) \cite{W01,K03}.

Since the first independently proposed KEDF by Thomas \cite{T01} and Fermi \cite{F01} (TF), a huge number of KEDFs has been suggested. However, \TF\ model with various corrections dominates the field \cite{X01,K01,K03}. The usual corrections are either based on gradient expansion \cite{ WOS:A1978EK01600046,WOS:000244532300067,WOS:A1981MP04500025} or on adding additional class of KEDF \cite{K01,W02,WOS:000258190400022,WOS:000077279800038}. Furthermore, \TF\--based KEDF are used in some applications satisfying the assumption that the density is nearly uniform. For example, they are used for metals \cite{H01} and warm dense matter \cite{W03,S02}. As for the correction, it would be assumed abstractly that gradient expansion should pave a reasonable route. However, it is known that the high order gradient-based corrections diverges for finite systems. Therefore, seeking non-gradient-expansion based correction is desirable. One of the recent corrections  suggested by Burke and coworkers \cite{ribeiro2015}  is based on uniform WKB analysis in 1-dimensional case. The addition has neither sums nor derivatives.

In this paper, we present a new non-gradient based method to enhance the \TF\ approximation by adding a correction factor derived using an exactly solvable model. The used model is for $N$ non-interacting particles moving in the Coulomb\st{ic} field of the nuclear charge. This allows using the simple Rydberg formula to calculate part of the energy and the particle density is expressed in an analytic form through Laguerre polynomials and exponential functions. The correction exploits the fact that the difference between the \TF\ approximation and the non-interacting kinetic energy is comparable to the difference between the same values inside the proposed solvable model. The concept can be presented best using the following equation
\begin{equation}
T_s= T_\mathrm{TF}+ (T_\mathrm{s}-T_\mathrm{TF})\approx T_\mathrm{TF}+ (\tilde{T}-\tilde{T}_\mathrm{TF}).
\label{tsappr}
\end{equation}
where $T_s$ is the non-interacting kinetic energy and $T_{TF}$ is \TF\ approximation. In the same equation we use the notation $\tilde{T}$ and $\tilde{T}_{TF}$  for the kinetic energy and Thomas-Fermi energy of the proposed model.
Our work follows a similar approach that was developed to estimate the correlation energy for two-electron atoms \cite{WOS:A1993LJ30800044}. In our numerical experiments, we show that the correction factor can  increase the precision of \TF\ approximation by around an order of magnitude for atoms.
 To apply it for molecules, the method needs further extensions. As shown in \eref{tsappr}, the kinetic energy is expressed as \TF\ energy plus the correction obtained from the exactly solvable model. We calculated the results numerically for a wide range of  atoms up to xenon ($Z=54$). Our results appear to be more accurate in comparison with \TF\ approximation by a factor between 11 (for helium) and 72 for xenon. We also discuss the large-$Z$ limit of the energy and the density and compare the correction to those by gradient expansions.

\section{The correction}
\label{Correction}

As aforementioned, the general idea of the proposed correction is that the difference between the \TF\ approximation and the non-interacting kinetic energy is comparable to the difference between the same values inside the proposed solvable model as shown in \eref{tsappr}. This follows a successful and analogous method used to estimate the correlation energy for two-electron atoms\cite{WOS:A1993LJ30800044} . Although we assume the model of an $N$ non-interacting particles moving in the Coulombic field of the nuclear charge, the method can be applied for a wide range of potential models. In this paper, the focus is to present the approach and to illustrate its applicability for simple systems like atoms. So, we suggest the use of a model that in many respects resemble the atoms and that has an analytical quantum-mechanical solution. Another important reasons for selecting such a model is that  it is possible to write the correction $\tilde{T}-\tilde{T}_\mathrm{TF}$ as a function of only the electrical charge $Z$. Furthermore, this form for $\tilde{T}$ and $\tilde{T}_\mathrm{TF}$ provides  more in depth understanding of the the model as it will be shown.

Since we are interested in an equivalent system of non-interacting particles, the energy $\tilde{T}$ does not include correlation effects. The inter-electron interaction affects the effective potential and therefore the only difference between our model and an atom is disregarding the screening of nuclear charge by inner electrons. The presented model retains Coulomb singularity at the origin, as well as Coulomb attraction for large distance, however the attraction force at large distance is much larger in the presented model. The region of applicability of the \TF\ model, $dp_{\mu}^{-1}/dr\ll 1$, where $p_{\mu}= \sqrt{2(\mu-V)}$ and $\mu$ is the chemical potential, is violated in a small region of radius $r_0=Z^{-1}$ adjoining the nucleus, where quantum effects become significant. In the neighborhood of the nucleus, the field is practically identical to the Coulomb field $-Z/r$. In our approach however, we treat the Coulomb problem exactly, without considering the quantum corrections separately.

In the assumed model, we consider a system with the nuclear charge $Z=N$.  Thus, the screened Coulomb potential for the atom has the same behavior $-Z/r$ at the origin, but differs far from it.  Without inter-particle interaction, each electron can be considered as occupying an orbital characterized by the principal quantum number $n=1,2,\ldots$. For example, two electrons with $n=1$ in a configuration $1s^2$ form the closest inner K-shell (in X-ray notations), eight electrons with $n=2$ form L-shell in a configuration $2s^2\,2p^6$. Generally, a completely filled $n$-th shell has $2 n^2$ electrons with possible quantum numbers $l=0,1, \ldots, n-1$, $m=-l,-l+1, \ldots, l$, and spins $\sigma=\pm\frac12$.

Let us denote $n_\mathrm{max}$ as the last shell with a non-zero occupation number. In the ground state, all lowest shells with $n< n_\mathrm{max}$ are completely filled, while the last shell with $n=n_\mathrm{max}$ can be filled either partially or completely. For simplicity, we consider initially only the states in closed-shell configurations, i.e. when all shells up to $n=n_\mathrm{max}$ are completely filled. The same model can be extended for other configurations either by considering angular dependence or simply by interpolation between closed shells as it will be shown later. The total number of electrons on shells with $n=1,2,\ldots,n_\mathrm{max}$ can be obtained by summation of the occupation numbers for each individual shell,
\begin{equation}
N= \sum_{n=1}^{n_\mathrm{max}} 2n^2= \frac13 n_\mathrm{max}(n_\mathrm{max}+1)(2n_\mathrm{max}+1).
\label{nel}
\end{equation}

The kinetic energy of the proposed model can be calculated by exploiting the kinetic energies for individual orbitals. This energy for each orbital is given by Rydberg formula, that is $Z^2/(2n^2)$ in atomic units, where $n$ is the principal quantum number.   Now, the total kinetic energy can be calculated by summation of contributions from each shell,
\begin{equation}
\tilde{T}= \sum_{n=1}^{n_\mathrm{max}} 2n^2 \left(\frac{Z^2}{2n^2}\right)= n_\mathrm{max} Z^2.
\label{ttotal}
\end{equation}
where the kinetic energy of the model $\tilde{T}$ corresponds to the same value in \eref{tsappr}.

To fully specify the proposed functional, we also need to calculate the Thomas-Fermi approximation of energy for the model. As previously stated, the  $\tilde{T}_{TF}$ can be presented as a function of the  electrical charge $Z$ when the proposed model is used. For closed-shell configuration which are spherically symmetric, \TF\ kinetic energy is
\begin{equation}
T_\mathrm{TF}= 4 \pi \int_{r=0}^{\infty} r^2\tau_0(r)\,dr,
\label{etf}
\end{equation}
where the KEDF $\tau_0$ is
\begin{equation}
\tau_0(\vec{r}) =\frac{3}{10} \left(3 \pi^2\right)^{2/3} \rho^{5/3}(\vec{r}).
\label{tau0}
\end{equation}

To be able to calculate the Thomas-Fermi approximation using \eref{etf}, for the proposed model, it is necessary to  have the corresponding electron density $\tilde{\rho}$. From the definition of the model, we know that $\tilde{\rho}$ can be acquired  by  combining the wave functions for individual electrons, as follows. A wavefunction of an electron on the orbital $(n,l,m)$ is
\begin{equation}
\psi_{n,l,m}(\vec{r})= R_{n,l}(r) Y_{l,m}(\theta,\phi),
\label{psinlm}
\end{equation}
where $\psi$ is a wave function, $n$ is the principal quantum number, $l$ is an orbital quantum number, $m$ is an azimuthal quantum number, $r$ is the radius, $Y_{l,m}$ is a spherical harmonic, $\theta$ is a polar angle, $\phi$ is an azimuthal angle. In \eref{psinlm}, $R_{n,l}$ represents the  the radial component of the wavefunction and it is defined using the following equation
\begin{equation}
R_{n,l}(r)= \sqrt{\left(\frac{2Z}{n}\right)^3 \frac{(n-l-1)!}{2n(n+l)!}}\exp\left(-\frac{Zr}{n}\right) \left(\frac{2Zr}{n}\right)^l
L_{n-l-1}^{2l+1}\left(\frac{2Zr}{n}\right).
\label{rnl}
\end{equation}
Using \eref{psinlm} the electronic density of an atom $\tilde{\rho}$ in our model  can be given as
\begin{equation}
\tilde{\rho}(\vec{r})= \sum_{n=1}^{n_\mathrm{max}} \sum_{l=0}^{n-1} \sum_{m=-l}^{l} R_{n,l}^2(r) |Y_{l,m}|^2(\theta,\phi).
\label{rhon}
\end{equation}
By exploiting the properties of the spherical harmonics and by summing over $m$, the above equation is reduced to
\begin{equation}
\tilde{\rho}(r)= \frac1{4\pi}\sum_{n=1}^{n_\mathrm{max}} \sum_{l=0}^{n-1} (2l+1) R_{n,l}^2(r).
\label{rho}
\end{equation}
which is angular independent. Finally we can calculate the value of $\tilde{T}_{TF}$ by incorporating  $\tilde{\rho}$, given by \eeref{rho}{tau0}.

After determining the values for $\tilde{T}$ and $\tilde{T}_{TF}$, we can explicitly write the functional for calculating the kinetic energy for an atom using the correction based on the proposed model. The new functional is given by the following equation
\begin{equation}
T\approx T_\mathrm{TF}+ \delta \tilde{T},
\label{approx}
\end{equation}
where $\delta \tilde{T}$ is found by considering the exactly solvable model with the same nuclear charge $Z$,
\begin{equation}
\delta \tilde{T}= \tilde{T}- \tilde{T}_\mathrm{TF}.
\label{deltat}
\end{equation}

When using the proposed functional for the kinetic energy of an atom, we would  calculate the $T_{TF}$ based on some electron density approximation and  the correction using the value of $Z$. In practical application,   it is necessary   to consider the case when the last shell is partially filled. In such cases, it is not possible to directly apply \eref{nel} with an integer $n_\mathrm{max}$. Since the closed shells occur only for a few values of the nuclear charge given by the sequence of ``magic numbers'' $2,10,28,60,110,\ldots$, we need to define the interpolation of the function $Z\mapsto \delta T'$ to other integer values of $Z$. Here, we use an interpolation by a cubic polynomial based on four points $Z=2$, $10$, $28$, and $60$:
\begin{equation}
\delta \tilde{T}= 0.21210 - 0.19860 Z +
 0.12815 Z^2 + 0.00010 Z^3.
\label{deltat3}
\end{equation}

\section{The $Z$ expansion of the model}

For many applications, it is more convenient to represent the energy of the system based on the total nuclear charge $Z$. This form can be acquired by solving \eref{nel} in respect to $n_\mathrm{max}$ and substitution of the result into \eref{ttotal}, we obtain the energy as a function of the nuclear charge,
\begin{equation}
\tilde{T}= \frac12\left(3^{-1/3}D^{-1}+ 3^{-2/3}D- 1\right) Z^2,\quad D=\left(54 Z+\sqrt{2916 Z^2-3}\right)^{1/3}.
\label{ttotalz}
\end{equation}
A more suitable format for \eref{ttotalz} is in the form of an expansion in powers of $Z^{-1/3}$. The transformed equation has the following form
\begin{eqnarray}
\tilde{T}   &\sim (3/2)^{1/3} Z^{7/3}  - \frac12 Z^2+ \frac1{6\times 12^{1/3}} Z^{5/3}-
\nonumber
 \frac1{3888\times 18^{1/3}} Z^{1/3} \\&+
 \frac1{69\,984\times 12^{1/3}} Z^{-1/3}+ O(Z^{-5/3}),
\label{ekinz}
\end{eqnarray}
where the terms proportional to $Z^{\pm 4/3}$, $Z^{\pm 1}$, $Z^{\pm 2/3}$, and $Z^{0}$ are identically zero. In numerics, \eref{ekinz} is simply
\begin{eqnarray}
\nonumber
\tilde{T}&\sim  1.144714 Z^{7/3}- 0.5 Z^2+ 0.072798 Z^{5/3}\\
&- 0.000098 Z^{1/3}+ 0.000006 Z^{-1/3}+ O(Z^{-5/3}).
\label{ekinzn}
\end{eqnarray}

With the goal of being able to assess the validity of the model for calculating the kinetic energy, we compare it to the corresponding expansion of Thomas-Fermi approximation, where the energy of an atom with large nuclear charge $Z$ can be written as an asymptotic series in powers of a small parameter $Z^{-1/3}$. This expansion is found to be \cite{schw1985},
\begin{equation}
T\sim 0.768745 Z^{7/3}- 0.5 Z^2+ 0.269900 Z^{5/3},
\label{ekinzat}
\end{equation}
The largest absolute difference between \eeref{ekinz}{ekinzat} is the leading term.
This is mainly due to the inclusion of Coulomb repulsion between electrons which decreases the \TF\ energy by a factor of $1.489$ because of the increase of the size of the atom. The second subdominant term $\sim Z^2$ remains the same. This term comprises the correction of strongly bound electrons \cite{scott1952,schw1980} which is not affected by the inter-electron repulsion.

To assess the accuracy of the \TF\ approximation, we analyze the asymptotic behavior of the \TF\ energy in the limit of large number of electrons $N=Z$. To estimate the large-$N$ behavior, we have calculated $\tilde{T}_\mathrm{TF}$ for increasing values of $n_\mathrm{max}$ and found, using Richardson's extrapolation \cite{bender}, that for large $n_\mathrm{max}$
\begin{equation}
\tilde{T}_\mathrm{TF}\sim 1.144714 Z^{7/3}- 0.625856 Z^2+ 0.146878Z^{5/3}.
\label{tfnexp}
\end{equation}
We characterize the accuracy of the approximation by measuring the deviation of the expansion (\ref{tfnexp}) from the exact coefficients in \eref{ekinzat}\cite{WOS:000279144500002}. We find that the \TF\ approximation correctly reproduces the leading term $1.144714 Z^{7/3}$, but makes a $25\%$ error in the subdominant term $-0.5 Z^2$. Thus we can say according to the definition from\cite{elliott} that the \TF\ approximation is large-$N$ asymptotically exact to the first degree (AE1).

\section{Numerical results and discussion}

In this section, the presented functional and the corresponding approach are implemented for verification. First we compare the electron density of the proposed model and its large-$Z$ limit.  Then, we compare the model to the known gradient expansion corrections of \TF\ KEDF. Finally, we perform computational experiments to show that the proposed functional gives a significant improvement to the standard Thomas-Fermi approximation when applied to atoms. An improvement by about an order of magnitude is achieved.

\subsection{Electron density in the limit of large $Z$}

\label{sectf}

In this subsection, we analyze the behavior of electron density of the proposed model to the one acquired using the Thomas-Fermi formalism vs. $Z$. We start with some remarks regarding the density of the model. In Section \ref{Correction}, the electron density is derived from the wavefunction as a sum of $n_\mathrm{max}(n_\mathrm{max}+1)/2$ terms given by \eref{rho}. For large $Z$, the number of terms grows to infinity. Here, we use an alternative approach based on \TF\ formalism to derive the  limit of the density for large values of $Z$ in a more explicit way. Without an inter-particle interaction, the \TF\ equation relating the electron density and the potential takes especially simple form\cite{WOS:A1981ML75100003},
\begin{equation}
\rho = \frac1{3\pi^2}\left[2(\mu-V)\right]^{3/2},
\label{rhotf}
\end{equation}
where $\mu$ can be determined from the equation (assuming spherical symmetry)
\begin{equation}
\int_0^{r_\mathrm{m}}4\pi r^2\rho(r)\,dr= N.
\label{mueq}
\end{equation}
In \eref{mueq}, $N$ is the number of electrons and $r_\mathrm{m}$ is the \TF\ radius of the atom, or a turning point in the potential $V$, that is determined from the equation
\begin{equation}
V(r_\mathrm{m})= \mu.
\label{rmeq}
\end{equation}
For the Coulomb potential $V= -Z/r$, we obtain
\begin{equation}
r_\mathrm{m}=-\frac{Z}{\mu}
\label{rmeq1}
\end{equation}
From \eeref{mueq}{rmeq1}, it can be shown that
\begin{equation}
r_\mathrm{m}= Z^{-1}\left(3\sqrt{2}N\right)^{2/3}.
\label{rmeq2}
\end{equation}

As we consider the the case of a neutral atom ($N=Z$), then, the dependence on $Z$ can be eliminated by introducing scaled radius and scaled electron density
\begin{equation}
\hat{r}= Z^{1/3} r,
\label{rsc}
\end{equation}

\begin{equation}
\hat{\rho}= \rho/Z^2,
\label{rhosc}
\end{equation}
so that now \eref{rhotf} can be rewritten as
\begin{equation}
\hat{\rho}= \frac{2\sqrt{2}}{3 \pi^2} \left(\frac1{\hat{r}}- 18^{-1/3}\right)^{3/2}
\label{rhotfsc}
\end{equation}
if $\hat{r}<\hat{r}_\mathrm{m}$ and zero otherwise, where the scaled turning point is
\begin{equation}
\hat{r}_\mathrm{m}= 18^{1/3}.
\label{rsc}
\end{equation}

Figure \ref{fig1} shows the scaled densities $\hat{\rho}(\hat{r})= Z^{-2} \tilde{\rho}(Z^{1/3}r)$, where $\tilde{\rho}$ is given by \eref{rhon} for increasing numbers of electronic shells, $n_\mathrm{max}=1,2,3,5$, calculated using \eref{rhon} together with the limiting case given by \eref{rhotfsc}. It is clear that the deviation from the \TF\ limit, $\tilde{\rho}-\tilde{\rho}_\mathrm{TF}$ is an oscillating function having exactly $n_\mathrm{max}$ local maxima, see the insert in Figure \ref{fig1}.  The amplitude of the oscillations decrease as $Z$ increases.  These oscillation effects are related to the shell structure, with maxima corresponding to the filled shells. Clearly, the \TF\ model describes only the averaged physical quantities, and so it requires a special generalization \cite{kirzhnits1972} to treat such spatial irregularities. In the recent paper of Burke and coworkers \cite{ribeiro2015}, their approximation improves the accuracy everywhere including the turning points and reproduces the oscillations.
\fig{Scaled densities for increasing values of $Z$ and \TF\ limit at $Z\rightarrow \infty$.  } {fig1} {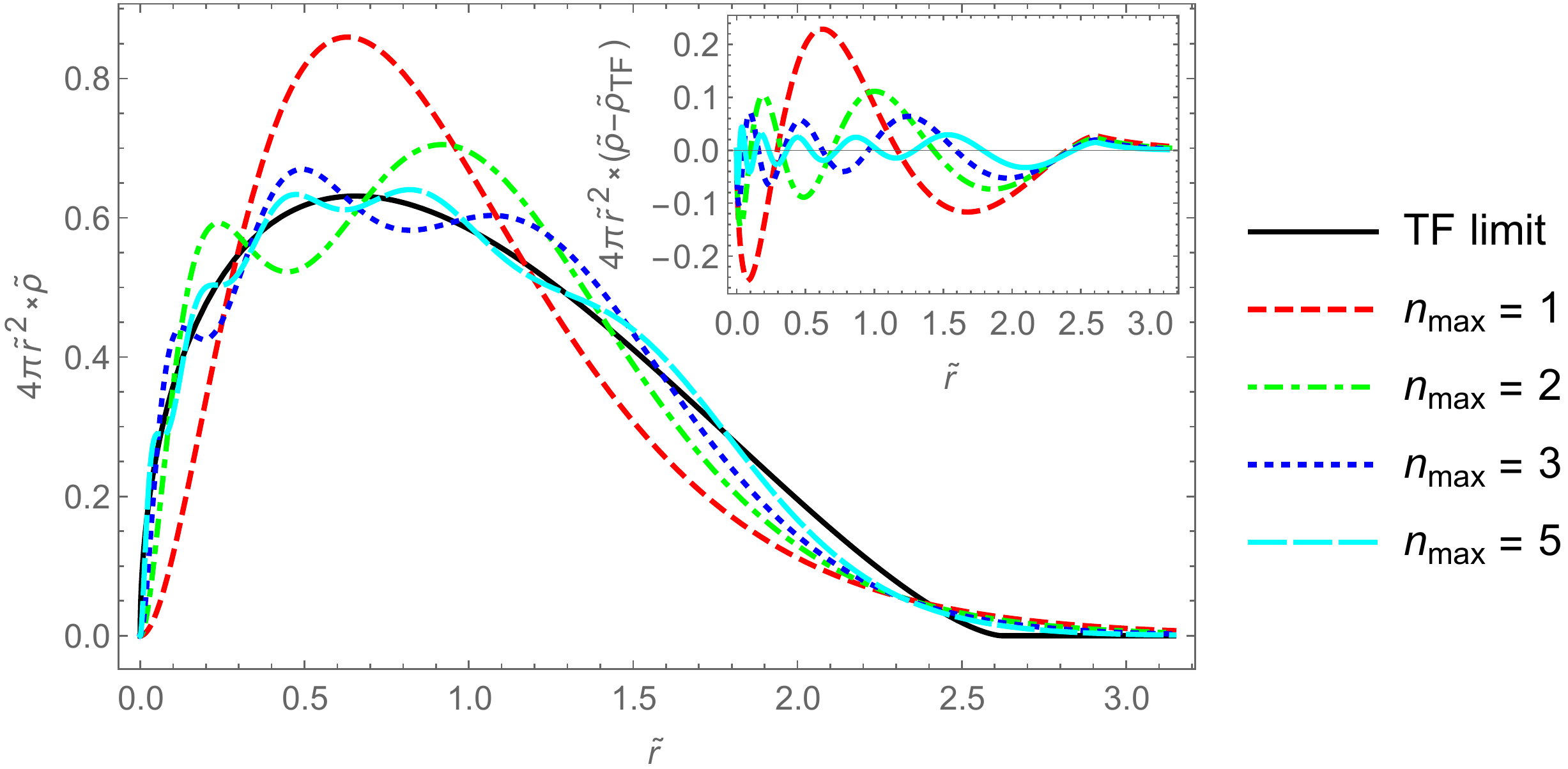}

\label{secdens}

\subsection{\TF\ vs. \GE\ for the exactly solvable model}
\label{secge}

The leading term of \GE\ for the kinetic energy is just the \TF\ energy given by \eeref{etf}{tau0}. We use the notation $T_0= T^{(0)}= T_\mathrm{TF}$. The second order term is defined through \Wz\ correction $T_\mathrm{W}$ as
\begin{equation}
T_2= \frac19 T_\mathrm{W},\quad T_\mathrm{W}= 4 \pi \int_{r=0}^{\infty}  r^2 \frac{(\rho')^2}{8\rho}\,dr,
\label{en2}
\end{equation}
and the second order approximation is $T^{(2)}= T_0+ T_2$, where spherical symmetry is assumed. In fourth order of the \GE, we have
\begin{equation}
T_4= 4 \pi \int_{r=0}^{\infty}r^2\, \tau_4(r)\,dr, \quad T^{(4)}= T_0+ T_2+ T_4,
\label{en4}
\end{equation}
where
\begin{equation}
\tau_4(\vec{r}) =
\frac{(3\pi^2)^{-2/3}}{540} \rho^{1/3}
\left[\frac{\left(2\frac{\rho'}{r}+\rho^{\prime\prime}\right)^2}{\rho^2}-
\frac98 \frac{\left(2\frac{\rho'}{r}+\rho^{\prime\prime}\right)(\rho')2}{\rho^3}+
\frac13 \frac{(\rho')^4}{\rho^4}\right].
\label{tau4}
\end{equation}

There were several studies of the \GE\ for atoms\cite{WOS:A1980JB99900060,WOS:A1980KB12700058,WOS:A1976CL12600003}. The \TF\ method always underestimates the energy, and the accuracy slowly improves with increase of number of electrons, remaining on the level of few percent even for heavy atoms. The first correction of \GE\ always improves the accuracy, but applying the fourth order correction of \GE\ typically makes the results worse. The analysis of trends of the \GE\ for atoms remains somehow inconclusive, because of very slow asymptotical behavior (typically as $\sim Z^{-1/3}$) and because of shell effects (oscillations of density).

\begin{figure}[tcb]
\centering
\includegraphics[width=0.85\textwidth]{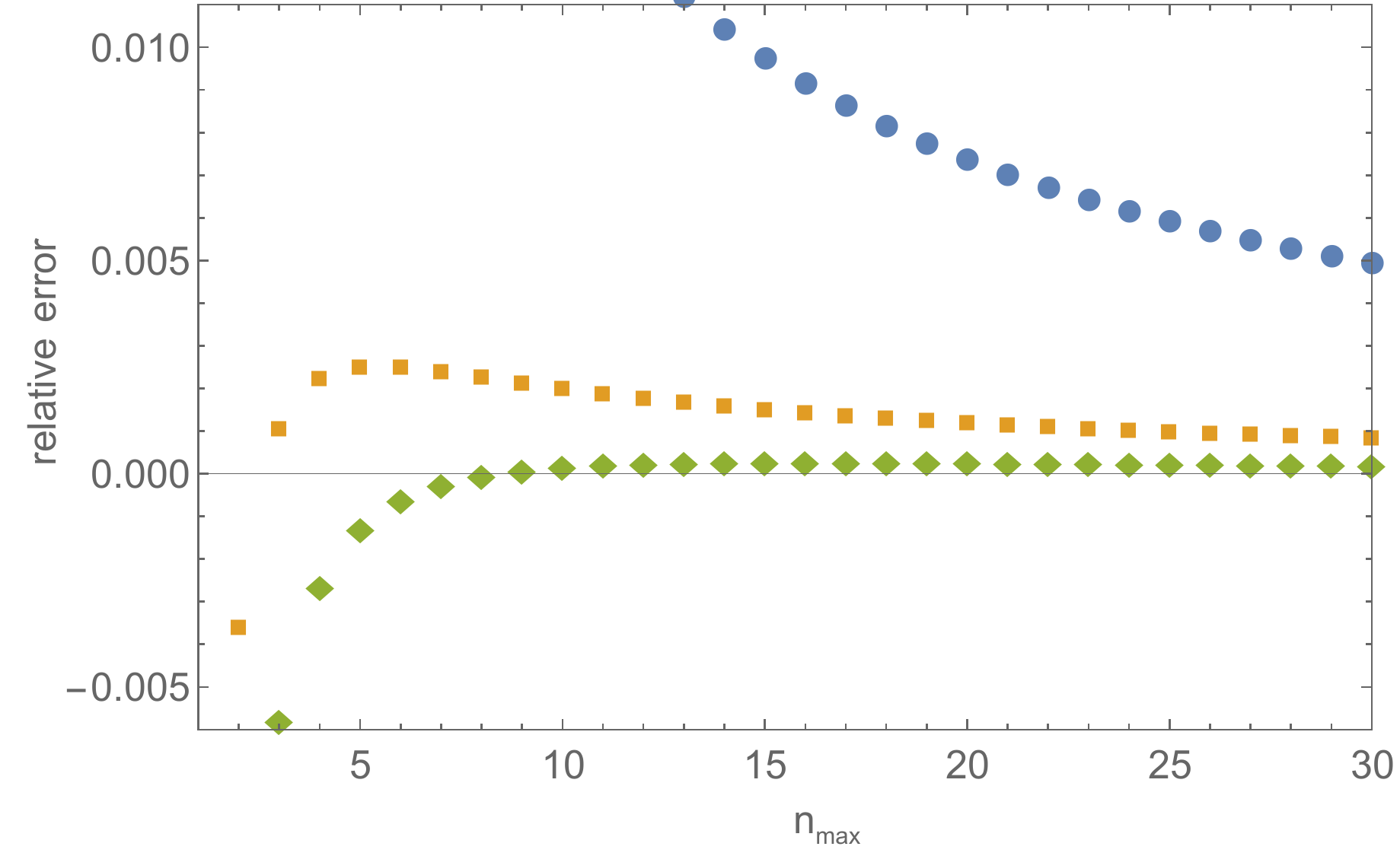}
\caption{Relative error of $T_\mathrm{s}$ for \TF\ ($T^{(0)}= T_\mathrm{TF}$, marked by circles), second ($T^{(2)}= T_0+T_2$, square markers), and fourth order approximation ($T^{(4)}= T_0+T_2+T_4$, diamond markers) in \GE . The relative errors are defined as $(T_\mathrm{s}-T^{(n)})/T_\mathrm{s}$ where $n=0,2,4$.}
\label{fig1a}
\end{figure}

For the exactly solvable model, relative error as a function of the number of shells is shown on Figure \ref{fig1a}, where the non-interacting kinetic energy $T_\mathrm{s}=\tilde{T}$ is defined by \eref{ttotal}, the \TF\ energy is defined by \eref{etf}, and the \GE\ is defined by \eeref{en2}{en4}. For the exactly solvable model, the relative error to $T_\mathrm{s}$ is shown on Figure \ref{fig1a} as a function of the number of shells. The values for $T^{(0)}, T^{(2)}, T^{(4)}$  are calculated by  substituting $\tilde{\rho}$ into the corresponding equations.
For large $Z$, including the second order correction clearly improves the accuracy by a factor of 6, and including the fourth order correction improves the accuracy by an additional factor of 3. However, this trend is visible only for large $n_\mathrm{max}$, starting from $n_\mathrm{max}\approx 10$ corresponding to unrealistic $Z>1000$. It explains the fact that the forth order correction for atoms improves the accuracy only for very heavy atoms.

\begin{figure}[tcb]
\centering
\includegraphics[width=0.85\textwidth]{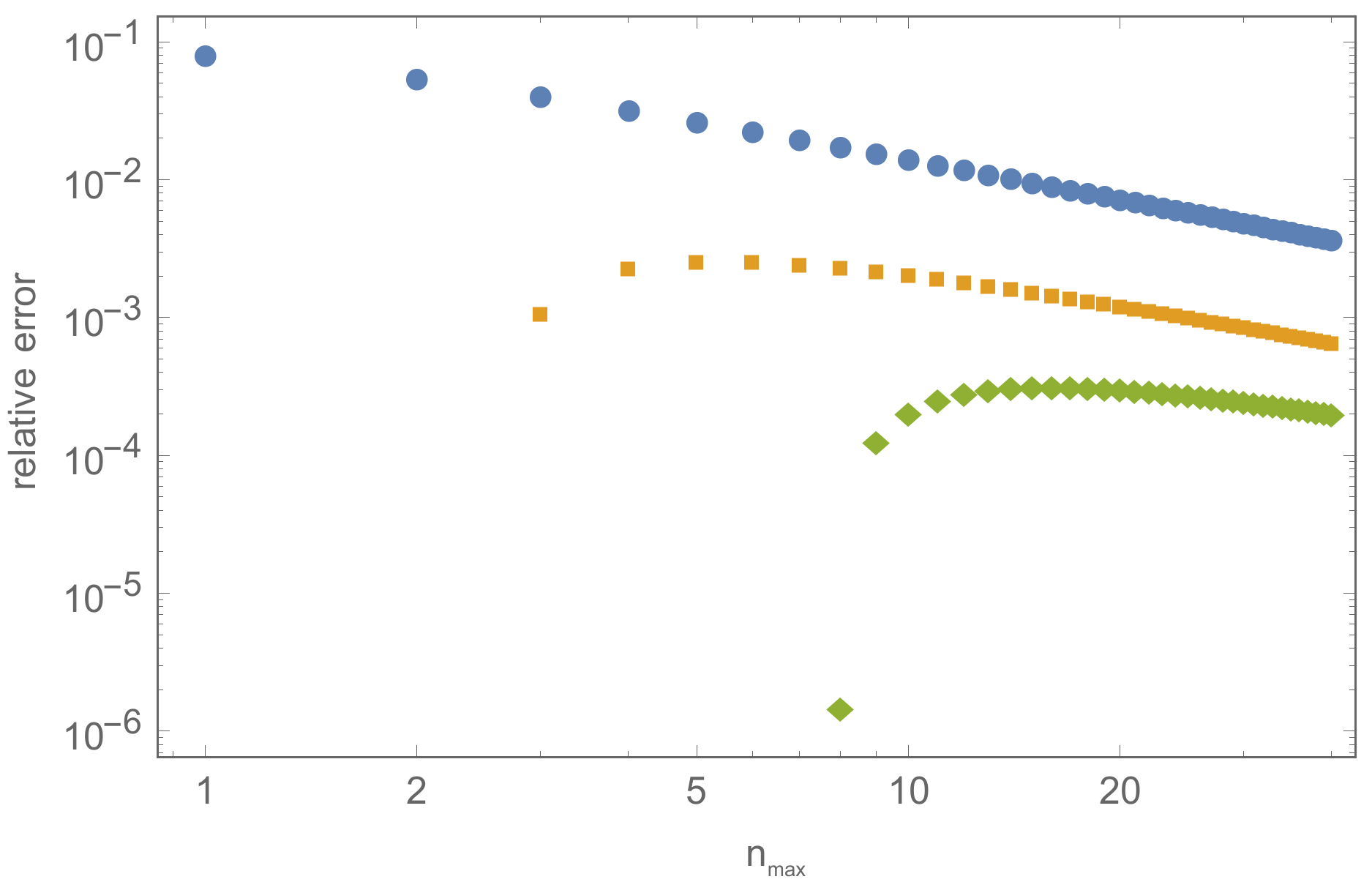}
\caption{Relative error in logarithmic scale for very large number of electronic shells for \TF\ approximation and the approximations including second and fourth order terms of \GE. The markers are the same as on Figure \ref{fig1a}. Since the asymptotic dependence is linear with the tangent (-1) in all cases, it shows that the relative errors at large $Z$ are proportional to $Z^{-1/3}$ in all cases.}
\label{fig2a}
\end{figure}

Asymptotic behavior of accuracy at large $Z$ is shown on Figure \ref{fig2a}.
The asymptotical expansion for the \TF\ energy  has been presented in the previous section and has the form given in \eref{tfnexp}. In a similar way the terms of \GE\ were determined numerically and have the following forms
\begin{equation}
T_2\sim  0.10942 Z^2+ 0.045 Z^{5/3}+ \ldots.
\label{t2exp}
\end{equation}
Notice that the expansion in \eref{t2exp} starts from the term $Z^2$, i.e. the coefficient of the leading term $\sim Z^{7/3}$ is zero. This statement can be proven rigorously too, by calculating the integral (\ref{en2}) using the same semiclassical approximation. A similar expansion (starting from the term $\sim Z^2$) was found for the fourth-order correction to the kinetic energy,
\begin{equation}
T_4\sim  0.015052 Z^2+ 0.0078 Z^{5/3}+ \ldots.
\label{t4exp}
\end{equation}

These calculation show that the leading term of the exact energy, $1.447 Z^{7/3}$, is absorbed by the leading term of \TF\ energy, and the subdominant term of the exact energy, $\sim -0.5 Z^2$, can be accurately resumed from the corresponding terms in the series $T^{(4)}= T_\mathrm{TF}+ T_2+ T_4$.

\subsection{Calculations for atoms}

\label{sectinterp}

In this subsection we discuss the effectiveness of the functional given by \eref{approx} for calculating the kinetic energy of atoms.  This is done by    calculating the kinetic energy for atoms using \eref{approx} and compare the results with \HF\ energies.
\label{secatoms}
To show explicitly the advantages of the proposed functional we compare the kinetic energies calculated using the new approach with the Thomas-Fermi approximation and the standard  functionals based on gradient expansion. We calculated \TF\ kinetic energy for atoms with closed shells using the electronic densities derived from Clementi Tables \cite{clementi}. The correction $\delta \tilde{T}$ was calculated by \eref{deltat} in case of closed shell atoms and by   \eref{deltat3} in partially filled cases. The results are shown in Table \ref{tab1}.
\setlength{\tabcolsep}{10pt}
\renewcommand{\arraystretch}{1.2}

\begin{table*}
\center
\caption{Comparison of \TF\ kinetic energy with the improved \TF, for atoms. Results from the second and fourth order gradient expansion are included too. The relative error is calculated in comparison with \HF\ kinetic energies taken from Clementi Tables \cite{clementi} which are known to be close to the energy $T_\mathrm{s}$. The density for atoms with nonzero angular momentum is spherically averaged.}

\begin{tabular}{|cccccc|}
\hline
\multicolumn{2}{|c}{ }& \multicolumn{4}{c|}{Relative error, \%} \\
\multicolumn{1}{|c}{$Z$}& \multicolumn{1}{c}{Atom}& \multicolumn{1}{c}{$T_\mathrm{TF}$}& \multicolumn{1}{c}{$T_\mathrm{TF}+T_2$}& \multicolumn{1}{c}{$T_\mathrm{TF}+T_2+T_4$}& \multicolumn{1}{c|}{$T_\mathrm{TF} +\delta T'$}\\[2pt]
\hline
2 & He& -11 &  0.59& 3.6& 0.95\\
3 & Li& -10 &  0.62& 3.1& 0.26\\
4 & Be& -9.9&  0.50& 2.9&0.21\\
5 & B & -10 & -0.53& 1.6&-0.52\\
6 & C & -11 & -1.3 &0.67&-1.0\\
7 & N & -11 & -1.7 &0.16&-1.2\\
8 & O & -10 & -1.6 &0.10&-0.95\\
9 & F & -9.4& -1.2 &0.37&-0.47\\
10&Ne&-8.4&-0.56&0.95&0.28\\
11&Na&-8.1&-0.49&0.94&0.37\\
12&Mg&-7.8&-0.44&0.95&0.43\\
13&Al&-7.6&-0.45&0.89&0.42\\
14&Si&-7.5&-0.47&0.83&0.39\\
15&P &-7.4&-0.50&0.77&0.36\\
16&S &-7.3&-0.51&0.72&0.35\\
17&Cl&-7.1&-0.51&0.70&0.35\\
18&Ar&-7.0&-0.49&0.69&0.36\\
36&Kr&-5.8&-0.69&0.18&0.11 \\
54&Xe&-5.2&-0.68&0.067&0.073 \\
\hline
\end{tabular}
\label{tab1}
\end{table*}

We found that including the correction $\delta \tilde{T}$ increases the accuracy of \TF\ approximation by more than  nine times.   As an approximation to the energy $T_\mathrm{s}$, here we used the \HF\ energies tabulated in \cite{clementi}.  The results are especially accurate for atoms with spin-paired electrons (He, Be, Ne, Mg, Ar, Kr, and Xe), where our numerical results always give the upper bound for the kinetic energy, while the \TF\ approximation gives lower bounds in all cases. We expect that spin-polarized version of \TF\ theory would work significantly better for cases with uncompensated spins, like atoms of B, C, N, O, F.

  Overall, the same table shows that the proposed method is competitive, with  functionals based on the  second and forth order gradient expansions when precision is considered. It manages to find approximations with a lower error than the second/forth order expansion  for 18/14 out of 19 tested atoms. From the tabulated results, it is  noticeable that the new functional is more robust than the other  functionals in the sense that it has a good performs  for both small and large atoms. It is important that the proposed correction procedure is much simpler to calculate, since it is a single variable function  compared to very complicated electron density functionals  in case of the gradient based corrections.

\section{Conclusion}
In this paper we have presented a non-gradient-based correction to \TF\ functional for atoms. The presented approach is general and should be extendible for molecules. The method uses an auxiliary system of noninteracting electrons that is in many respects similar to the  atomic system with the same number of electrons. It results in simplifying the calculations considerably when compared to the one based on gradient expansion. 
The obtained accuracy is improved by more than  nine times in comparison to \TF\ model. Our numerical test have also shown that the proposed method manages to achieve similar, slightly better, precision  that the standard gradient based functionals. As for the density, our presented approach allows the characteristic shell oscillations.

This type of approach can potentially be extended to systems other than atoms. One example is the modeling on $N$-electron quantum dots, where we could consider another solvable model, of $N$ non-interacting particles bound in a harmonic potential. Our initial test on this problem have given promising results.

\section{References}

\providecommand{\newblock}{}

\end{document}